\documentclass{emulateapj}
\usepackage{graphicx}
\usepackage[euler]{textgreek}
\usepackage{amssymb}
\usepackage{comment}
\usepackage{subfigure}
\usepackage{color}

\slugcomment{Accepted for publication in the \apj}

\shorttitle{Has AMS-02 observed anomalous cosmic rays?}
\shortauthors{Khiali et al.}

\begin{document}


\title{Anomalous Galactic cosmic rays in the framework of AMS-02}


\author{Behrouz Khiali \altaffilmark{1}, Sadakazu Haino\altaffilmark{2}, Jie Feng \altaffilmark{2,3}}

\altaffiltext{1}{National Central University (NCU), Chung–Li, Tao Yuan, 32054, Taiwan}
\altaffiltext{2}{Institute of Physics, Academia Sinica, Nankang, Taipei 11529, Taiwan}
\altaffiltext{3}{School of Physics, Sun Yat-Sen University, Guangzhou 510275, China}

\email{behrouz.khiali@cern.ch}

\begin{abstract}
The cosmic ray energy spectra of protons and helium nuclei, which are the most abundant components of cosmic radiation, exhibit a remarkable hardening at energies above one hundred GeV/nucleon. Recent data from AMS-02 confirms this feature with a higher significance. This data challenges the current models of cosmic ray acceleration in Galactic sources and propagation in the Galaxy. Here, we explain the observed break in the spectra of proton and helium nuclei in light of recent advances of cosmic ray diffusion theories in turbulent astrophysical sources as a result of a transition between different cosmic ray diffusion regimes. We reconstruct the observed cosmic ray spectra using the fact that transition from normal diffusion to superdiffusion changes the efficiency of particle acceleration and causes the change in the spectral index.  We find that calculated proton and helium spectra match very well with the  data.
\end{abstract}


\section{Introduction}\label{sec1}

Cosmic rays (CRs) carry a wealth of information on galactic astrophysics and possibly about new fundamental particle physics. Despite recent broad studies of galactic CR transport (e.g., \citealt{tomassetti12, aloisio15}), there are still difficulties to understand the spectral features of CRs, as well as their origins, acceleration mechanisms, and propagation. These long-standing problems are described in \cite{schlic02} and references therein. Recent spectral data from CR detectors (e.g., CREAM, PAMELA) has provided invaluable information for probing the properties of the interstellar medium, understanding Galactic magnetic fields and searching for dark matter (\citealt{bergstrom12}).

Data from the ATIC-2 (\citealt{panov09}), CREAM (\citealt{ahn10}) and PAMELA (\citealt{adriani11}) experiments  indicate that spectral shapes of the protons and helium nuclei cannot be explained by a single power law due to an observed  break in the rigidity range of a few hundred GV. This has been confirmed recently by the \textit{Alpha Magnetic Spectrometer}\footnote{AMS-02 is a space-borne high energy particle detector installed on the International Space Station (ISS). Its purpose is to perform accurate, high statistics, long duration measurements of the spectra of energetic (up to multi-TeV) primary charged cosmic rays (\citealt{ams13}). } (AMS-02) experiment (\citealt{aguilar15a,aguilar15b}). Current theoretical models have  explained these puzzling spectral features as  \textit{source effect} scenarios caused by different acceleration mechanisms (\citealt{biermann10}) or different populations of CR sources (\citealt{yuan11}). Moreover, there are other theoretical attempts to explain the observed upturn in the CR spectra based on the spatial change of CR diffusion properties of several sources in the Galaxy (see \citealt{tomassetti12,evoli14,aloisio15}).   

 In this paper, we propose that the observed upturn break in the CR spectra indicates a transition between different CR diffusion regimes close to the acceleration regions in supernova remnants\footnote{The Galactic CR spectrum in the energies above $\sim$10 GeV/nucleon is thought to be accelerated in SNRs (\citealt{tomassetti12}).} (SNRs), and that this transition substantially changes the efficiency of CR acceleration, causing a break in the power-law distribution of CRs.    
 
 Particle acceleration at termination shock in SNRs has been explained in the framework  of diffusive shock acceleration (DSA) based on first-order Fermi acceleration, proposed by \cite{fermi54}.  In this mechanism, depending on Larmor radius, particles bounce back and forth across the shock front due to their interactions with magnetic irregularities process and gain energy. However, DSA is not able to explain all energetic particle observations in astrophysical sources  and an alternative acceleration mechanism, namely \textit{magnetic reconnection} (see \citealt{zank15}), has been applied for some Galactic and extra-galactic sources to explain the CR acceleration (e.g., \citealt{khiali15a,khiali15b}).  

Magnetohydrodynamical (MHD) turbulence is ubiquitous in astrophysics and plays a key role in CR acceleration and diffusion. In the presence of turbulence, magnetic field lines diverge and CRs, which follow the magnetic field lines, are scattered by magnetic perturbations.

 Assuming a time evolution for the mean square displacement $\langle(\Delta x)^2\rangle$, we have 
\begin{equation}\label{displacement}
\langle(\Delta x)^2\rangle =2 D_{\sigma} t^{\sigma},
\end{equation} 
where $D_{\sigma}$ is a diffusion coefficient\footnote{\cite{zimbardo13} have derived an expression for  $D_{\sigma}$ in the superdiffusive regime of CR propagation.} with dimensions $[D_{\sigma}]=(\rm{length})^2/(\rm{time})^{\sigma}$. Using different values of $\sigma$, we can characterize the particle motion in different regimes. In the case $0<\sigma<1$, the particle motion is in subdiffusion regime; when $\sigma=1$, the normal (Markovian) diffusion dominates the particle motion; for $1<\sigma<2$ the regime of particle motion is superdiffusion; and in the case of $\sigma=2$, particles are moving ballistically or free streaming (\citealt{shalchi09book}).

In most cases in astrophysical turbulent plasmas, the particle motion falls into the normal diffusion ($\sigma=1$) regime, characterized by Gaussian statistics. However, it has been demonstrated that in a few cases, CR transport can be in subdiffusion or superdiffusion regimes which are denoted as anomalous diffusion ($\sigma \neq1$). Recently, \cite{lazarianyan} have demonstrated that the divergence of the magnetic field on scales less than the injection scale of the turbulence induces superdiffusion of cosmic rays (CRs) in the direction perpendicular to the mean magnetic field. The possibility of CR acceleration at interplanetary shocks in the regime of superdiffusive transport also  has been shown by \cite{perri07,perri08,perri09a,perri09b}. The effect of superdiffusion  has also been observed in numerical simulations by \cite{xu13} and \cite{roh16} for the particles accelerated by the shock mechanism.

It has been demonstrated that the anomalous diffusion modifies the energy spectral indices predicted by DSA. Particularly, in the case of subdiffusion, the particle spectrum is steeper than in the normal diffusion case (\citealt{kirk96}), while the spectral indices are smaller in the case of superdiffusion (\citealt{perri12,zimbardo13}). They showed that the particle spectral indices depend on compression ratio\footnote{The ratio between flow velocities before and after shock which is $1\le r \le 4$.} ($r$) and $\sigma$.

Applying this fact, that the transition from normal diffusion to superdiffusion changes the power law indices, we can explain the observed break in proton and helium nuclei spectra. We show that this model is consistent with AMS-02, PAMELA, and CREAM data.

The outline of this paper is as follows. In Section 2, we provide a brief description of the status of the observed break in proton and helium spectra. We discuss superdiffusive shock acceleration in Section 3. In section 4, we show  how the upturn in CR spectra reported by AMS-02, PAMELA and CREAM  can be explained using different particle diffusion regimes. Finally,  we discuss our results and draw conclusions in Section 5.

\section{Hardening in proton and helium spectra}
Galactic CRs with energies above a few GeV are assumed to be accelerated in supernova remnants (SNRs) based on  the pioneering proposal of \cite{baade34} with a source spectrum of $Q_S\propto E^{-\gamma}$, namely the spectra before propagation (at injection) in the Milky Way. However, the observed CR spectrum at Earth, namely the spectra after propagation in the interstellar medium (ISM), is steeper due to leakage from the Galaxy (\citealt{strong07}). 

The particle acceleration mechanisms are not fully understood yet. However, it is believed that in the ISM, the particles are accelerated by shock waves at SNRs by DSA and injected to the Galaxy with the spectra mentioned above with spectral index (e.g., \citealt{hillas05}):
\begin{equation}\label{dsa-index}
\gamma=\frac{r+2}{r-1},
\end{equation}
where $r={V_1}/{V_2}$ is the ratio between upstream ($V_1$) and downstream ($V_2$) plasma velocities in the shock frame which is called the \textit{compression ratio} of the shock. According to the shock theory, the maximum value for the compression ratio is $r=4$. The corresponding spectral index is $\gamma=2$, which represents the spectral index for the strongest shock in astrophysical sources. 

As stressed before, measured CR spectra at Earth are steepened by propagation in the ISM and require a diffusion coefficient $D(E)\propto E^{\delta}$ because interactions with ISM leads to production of secondary nuclei. Therefore,  the observed fluxes of primary particles (e.g., protons and helium nuclei) have  power-law spectra like $\sim E^{-\gamma -\delta}$ and primary to secondary ratios have spectra like  $\sim E^{-\delta}$ (e.g., the B/C spectrum with $\delta \sim 0.3-0.6$; \citealt{strong98,evoli08,blasi12}). The B/C ratio is a valuable source of information about CR propagation in the galaxy (see \citealt{busching05}).

Usually, the propagation of Galactic CRs is described by (\citealt{berezinskii90}):
\begin{equation}\label{N(E)}
\frac{\partial N(\textbf{r},p,t)}{\partial t}-\nabla (D_{xx}\nabla N)=Q(\textbf{r},p,t),
\end{equation}
where $N(\textbf{r},p,t)$ is the CR density per unit of total particle momentum $p$ at galactic position \textbf{r}. $D_{xx}$ is the spatial diffusion tensor. Cooling mechanisms and nuclear fragmentation are indicated by $Q(\textbf{r},p,t)$.  Obviously, from Eq.\ref{N(E)}, one can derive the observed CR spectra at the Earth as a single power-law $\propto E^{-\gamma - \delta}$ for energies $\gg 1$ GeV/nucleon, neglecting energy losses and nuclear interactions (\citealt{evoli14}).
\begin{figure}
 \centering
 \includegraphics[width=3.5in]{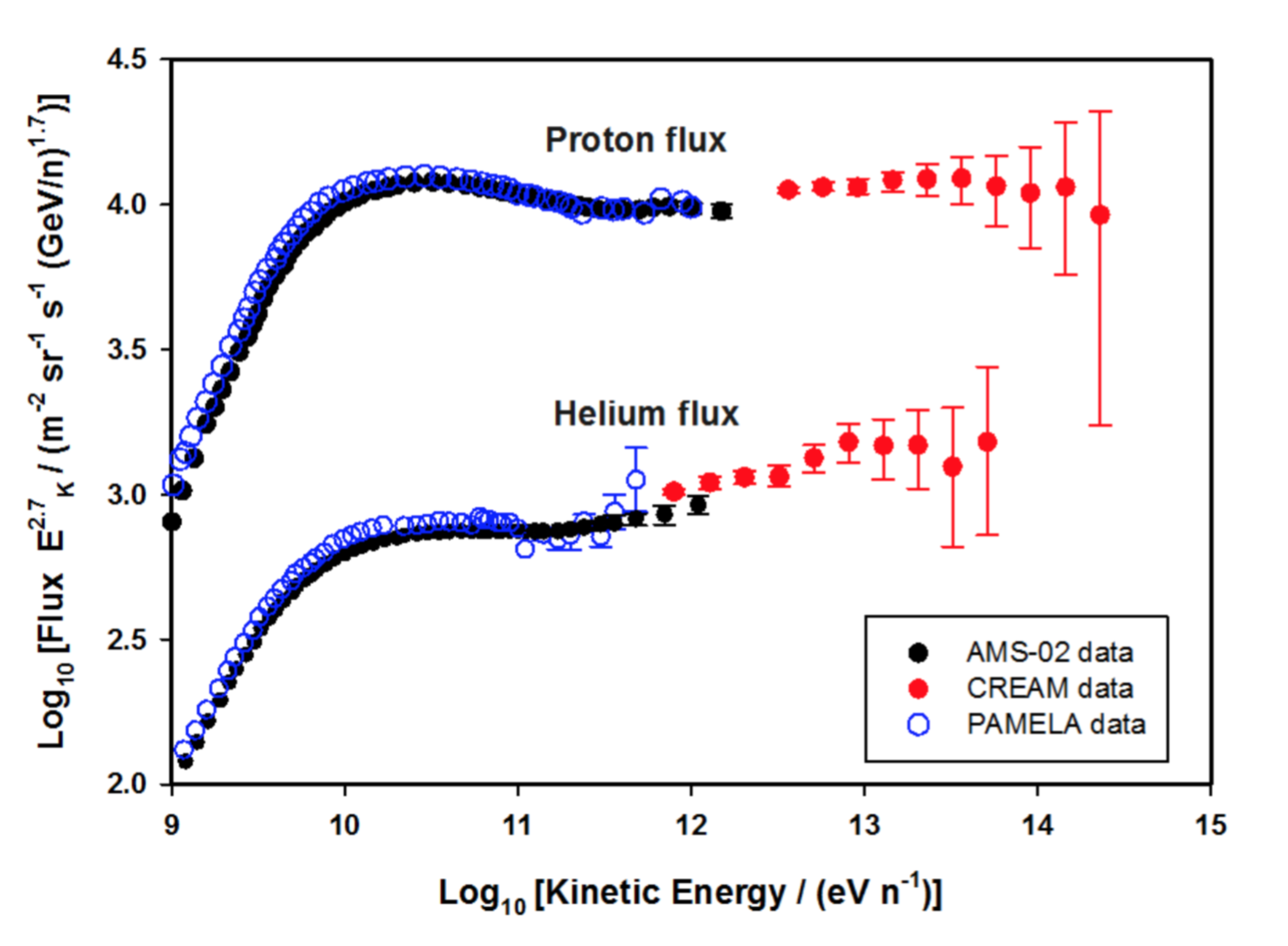}
 \caption{Proton and helium absolute fluxes measured by AMS-02, PAMELA and CREAM above 1 GeV per nucleon.}
 \label{data}
\end{figure}
However, the measurements by the PAMELA  experiment deviate from a single power-law, with a break in proton and helium nuclei spectra (see Fig.\ref{data}). In the observed proton spectrum, there is a change in slope at $\sim 230$ GeV,  from $\gamma + \delta=2.85$ to $\gamma + \delta=2.67$ for higher energies. The measurements of helium nuclei from PAMELA also exhibit a change in slope  at  $\sim 243$ GeV, from $\gamma + \delta=2.76$ to $\gamma + \delta=2.47$ (\citealt{adriani11}).

Recently, the accurate measurements of proton and helium nuclei by AMS-02 experiment (\citealt{aguilar15a,aguilar15b}) have improved on the results of previous experiments (see Fig.\ref{data}). The AMS-02 collaboration reported that the proton flux spectrum has a break at $\sim 330$ GeV, with $\gamma + \delta=2.85$ for $E<330$ GeV and  $\gamma + \delta=2.71$ for $330\ \rm{GeV}<E<\sim 2$ TeV. They have also observed a change in the slope of the helium flux at $\sim 245$ GeV.
According to data reported by AMS-02 collaboration, an upturn is observed for $E>245$ GeV, from $\gamma + \delta=2.78$ to $\gamma + \delta=2.66$.

In order to explain the high energy break in proton and helium nuclei spectra, we assumed that the diffusion coefficient can be approximated as a single power-law, using the value $\delta=0.35$ from B/C data with $D\propto E^{\delta}$.    This value is consistent with the new measurement of the AMS-02 experiment. Recently the AMS-02 collaboration reported their precise measurement of  B/C ratio above 65 GV, which  can be described by a single power law of  $\delta=0.33\pm 0.015$ (\citealt{aguilar16}). This value is in agreement with the Kolmogorov turbulence spectrum in the Galactic magnetic field, which predicts $\delta=1/3$ asymptotically (\citealt{kolmogorov41}). Using current B/C data, the uncertainty of $\delta$ is found to be ~5\%. 
So, in this experiment the measured values for $\gamma$ are as below: for the proton $\gamma=2.52 \pm 0.015$ for $E<330$ GeV and $\gamma=2.38 \pm 0.015$ for $E>330$ GeV and in the case of helium $\gamma=2.45 \pm 0.015$ for $E<245$ GeV and  $\gamma=2.33 \pm 0.015$ for $E>245$ GeV.

As discussed earlier, the value of $\gamma$ is related to CR injection from the source, and a smaller $\gamma$ means a more efficient acceleration mechanism injects the particles into the ISM. Thus, since the shock acceleration in SNRs will be more efficient in the presence of superdiffusion (\citealt{perri12}), these observed spectra tell us that we should expect to have a transition from DSA to superdiffusive shock acceleration. In the following sections, we further explain how to calculate the CR fluxes and demonstrate that the observed CR fluxes can be interpreted by the existence of different CR diffusion regimes in the SNRs.

\section{Superdiffusive shock acceleration model}

DSA is deemed as a proper particle accelerator at shock waves to describe how the particles can be accelerated in Galactic (e.g., SNRs) and extragalactic sources. However, there are some observational evidences to challenge this mechanism as the main acceleration process for CRs. For the strongest shocks ($r=4$), the standard spectral index has the value of 2 ($\gamma=2$, see Eq.\ref{dsa-index}), but some measurements point to harder spectral indices for CR spectra. For instance, observations of Crab Nebula in the range of radio emission indicate a spectral index for relativistic electrons with the value of $\gamma \sim 1.5$ (\citealt{hester08}) and recent precise measurements by AMS-02 (\citealt{aguilar15a,aguilar15b}), indicate that different spectral indices for proton and helium nuclei are inconsistent with the predictions of DSA, which only depends on shock compression ratio (Eq.\ref{dsa-index}).  

As stressed in \S.\ref{sec1},  recent simulations and theoretical approaches demonstrate the possibility of superdiffusive transport of particle at shocks (\citealt{lazarianyan}). They showed that fast deviations of magnetic field lines from the mean direction of the magnetic field cause superdiffusive transport  of CRs at shock waves and make a more efficient CR acceleration mechanism. This is due to substitution of the CR's mean free path for the length scale of \textit{magnetic field entanglement}\footnote{The entangled magnetic fields are naturally produced in the pre-shocked regions through the interaction of the precursor with density inhomogeneities in the ambient media or can be produced by various instabilities induced by CRs (\citealt{lazarianyan}).}.  These theoretical predictions have been tested numerically by \cite{xu13} in the presence of magnetic turbulence, and they have shown that on scales smaller than the turbulence injection scale, CR propagation is superdiffusive and the superdiffusive process is important for describing the propagation and acceleration of CRs in supernovae shells and shock regions. They found that the superdiffusive transport on small scales can also naturally explain the experimental data in the heliosphere. 

Moreover, the superdiffusive behavior of the solar energetic particles has been argued for based on analysis of particle time profiles (\citealt{perri09b}). They find that the propagation of energetic particles in interplanetary space is superdiffusive. There are plenty of numerical studies to probe the anomalous transport of CRs, both in directions parallel to the mean magnetic field (\citealt{zimbardo06,shalchi07}) and perpendicular to magnetic field lines (\citealt{shalchi09,ragot11,xu13}). 

\begin{figure}
 \centering
 \includegraphics[width=3.5in]{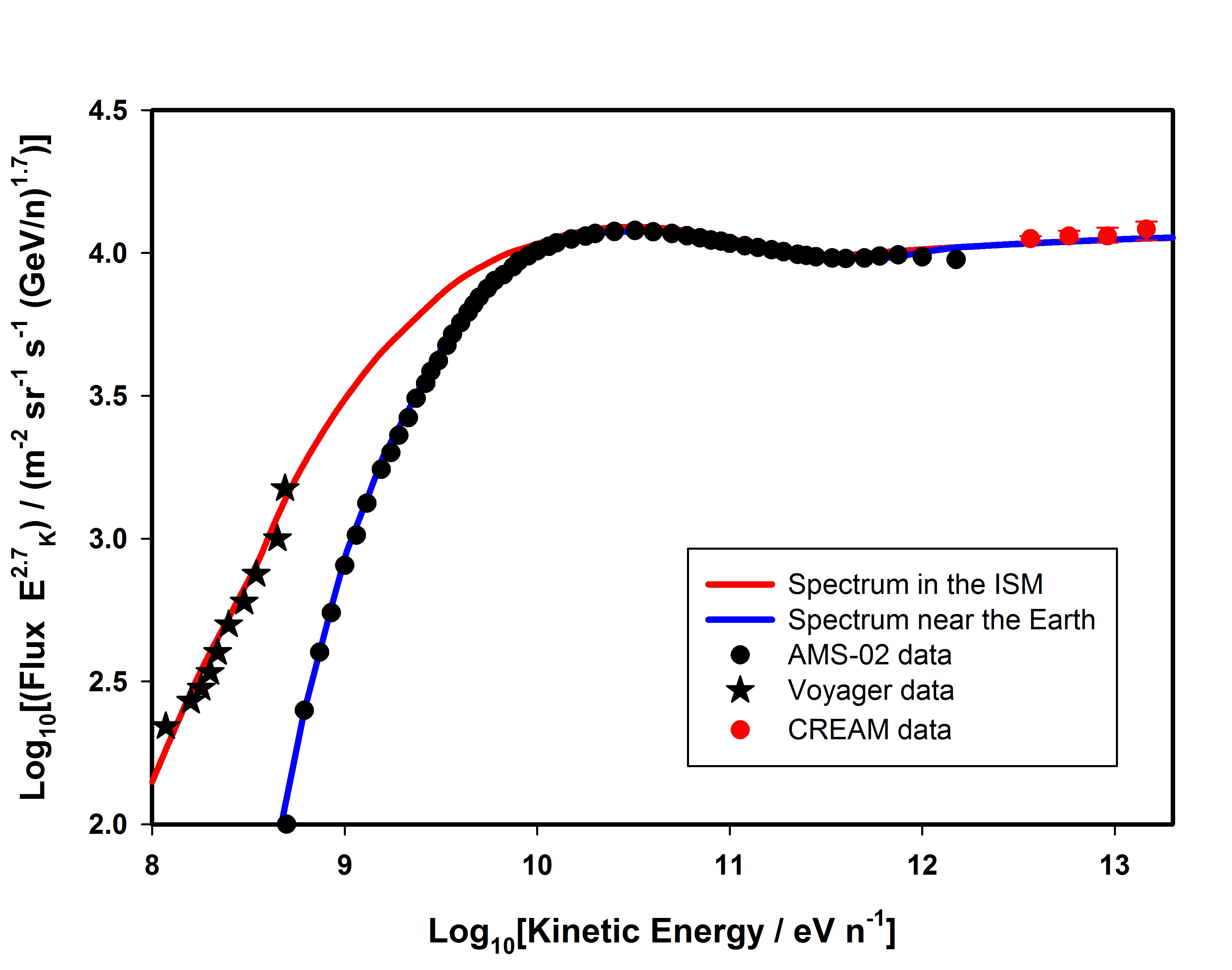}
 \caption{Spectrum of protons measured by Voyager (black stars,\citealt{stone13}), AMS-02 (black circles, \citealt{aguilar15a}) and CREAM (red circles, \citealt{yoon11}), compared with our calculations (lines). The red line is the spectrum in the ISM, while the blue line is the spectrum near the Earth after solar modulations.}
 \label{fig_proton}
\end{figure}

An alternative explanation for the energy spectral index of relativistic particles accelerated at shock fronts has been proposed by \cite{perri12} in the context of superdiffusive shock acceleration (hereafter SSA model). In this model, the values of spectral indices are smaller than the values predicted by DSA (see Eq.\ref{dsa-index}), making it possible to interpret  the observed spectral indices for those sources whose indices are smaller than $\gamma \sim 2$ and also for the relativistic accelerated particles in shocks which is given by (\citealt{perri12}): 
\begin{equation}\label{index-ssa}
\gamma=\frac{6}{r-1}\frac{2-\sigma}{3-\sigma}+1.
\end{equation}

As we see in Eq.\ref{index-ssa}, the spectral index in the framework of superdiffusive shock acceleration depends on both compression ratio ($r$) and the regime of diffusion ($\sigma$, see Eq.~\ref{displacement}). Thus, this model is able to solve the problems challenged by observations in DSA theory providing different spectral indices for any ion species which may explain different CR propagation regimes. SSA model has been applied to explain the properties of shocks and CR propagation for a number of interplanetary shock waves from spacecraft observations. 
For instance, \cite{perri15a} found that the acceleration times due to SSA are much shorter than in the DSA model, and shorter than the interplanetary shock lifetimes, as well.  

\begin{figure}
 \centering
 \includegraphics[width=3.5in]{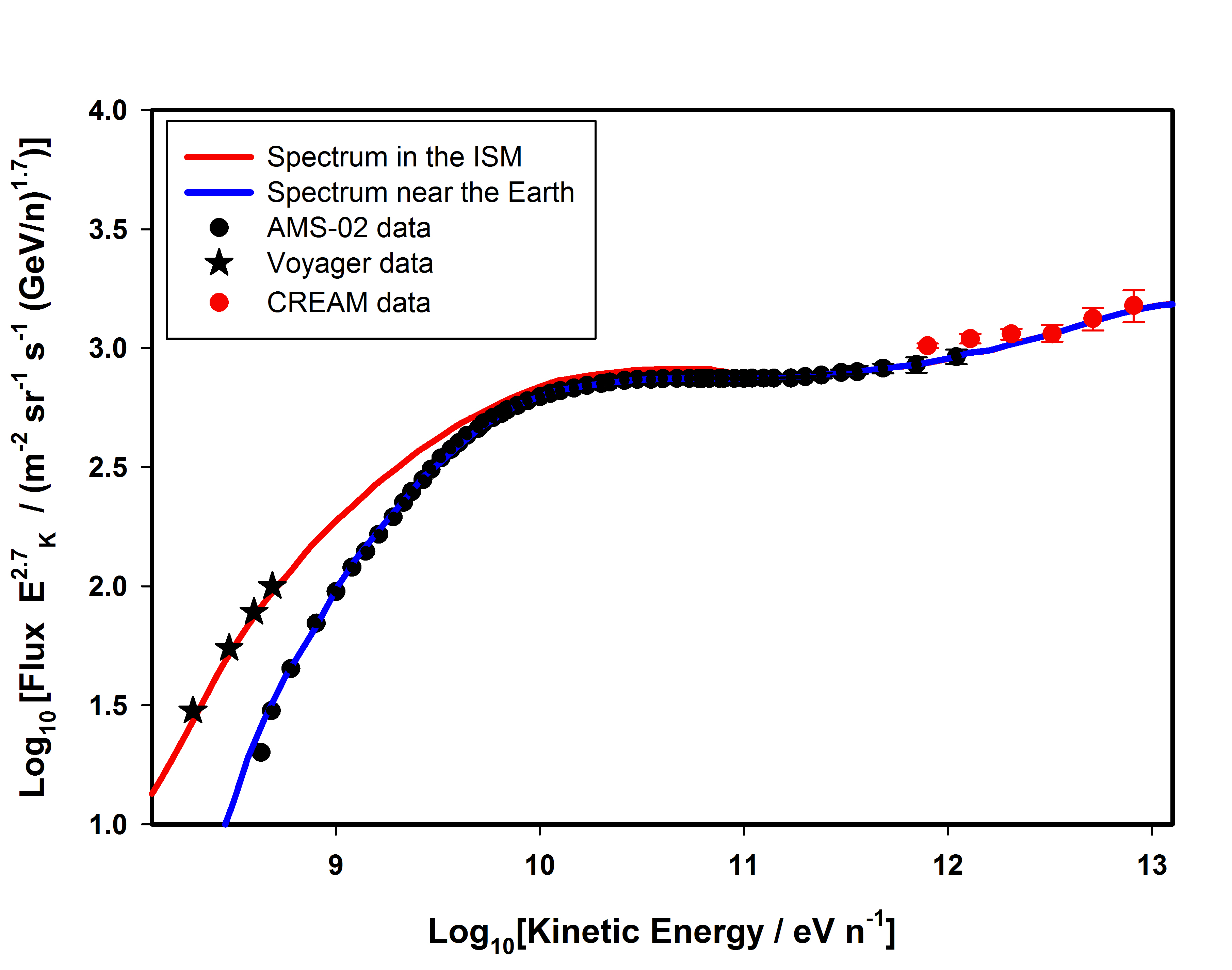}
 \caption{Spectrum of helium nuclei measured by Voyager (black stars,\citealt{stone13}), AMS-02 (black circles, \citealt{aguilar15b}) and CREAM (red circles, \citealt{yoon11}), compared with our calculations (lines). The red line is the spectrum in the ISM, while the blue line is the spectrum near the Earth after  solar modulations.}
 \label{fig_helium}
\end{figure}

Recently, it has been revealed by \cite{lazarianyan} and \cite{perri15a} that superdiffusive shock acceleration predicts new acceleration times, which are much shorter than in the DSA model, and also shorter than the interplanetary shock lifetime. The CR acceleration mechanism in the superdiffusive shock is faster because, in this regime, the particles have a larger displacement compare to DSA, as we see by the definition in Eq.\ref{displacement}. Therefore, this \textit{scale free-like} CR displacement increases the chance of particles far from shock front to return. In the vicinity of the shock front, possible short displacements of particles (with high probability) increase the chances of particles to cross the shock many times, gaining energy via first-order Fermi process faster in the superdiffusive shock acceleration (\citealt{perri15a}).  

As mentioned above, the predicted acceleration time in DSA theory is much longer than the values measured from observations. For instance, assuming the mean free path is equal to particle Larmor radius, it has been shown by \cite{lagage83} that the maximum possible energy gained in the available lifetime at SNRs shocks is $\sim 10^{14}$ eV which is much smaller than the CR energy at the "knee" in the CR spectrum $\sim 5\times 10^{15}$ eV. For non-relativistic ions in the heliosphere (with energies of $\sim$ a few MeV), which are believed to undergo additional acceleration at interplanetary shocks, the predicted acceleration time is much longer than the shock lifetime. Thus, it seems that the  particles are accelerated by a faster shock acceleration mechanism than DSA. Consequently, SSA model is able to provide a solution to overcome the problems of DSA challenged by observations regarding the acceleration time.

\section{calculation and results}

The observed upturn in proton and helium nuclei spectra at $\sim 300$ Gev means that the particles in this range of energy could be injected with a more efficient CR accelerator in the Galaxy.  Recently, it has been demonstrated that particles can be injected by superdiffusive shock acceleration, causing them to gain energy faster than the diffusive shock acceleration and to be injected with smaller spectral index. We assumed a transition from DSA to SSA may occur in SNRs. Thus, according to our proposed model, the measured spectral index for the energies before hardening is related to the DSA, and for the higher energies, the index corresponds to SSA.

Using the data from AMS-02, which is more precise with smaller systematic errors than the other CR experiments for the $\gamma$ before hardening region, we find the shock compression ratio ($r$) from Eq.\ref{dsa-index} for protons and helium nuclei.  Then using the  measured values of $\gamma$ in the higher energies, we are able to find $\sigma$, the superdiffusion index for each particle species. Thus, using our model and observational data, we can find the superdiffusion index, acceleration time in SSA and superdiffusive coefficient ($D_{\sigma}$)\footnote{For the expressions of  acceleration time in SSA and $D_{\sigma}$, see  \cite{perri15a}.}. According to AMS-02 data, injected protons are accelerated by shock acceleration with the compression ratio $r=3.00  \pm 0.04$ and in the superdiffusive regime,  $\sigma=1.16$. However, in the case of helium nuclei, we found that the compression ratio is $r=3.10 \pm 0.04$ and  $\sigma=1.15$, which means that the CR helium is produced in the stronger shocks.

\begin{table}[!ht]
\centering
\begin{minipage}{86mm}
\caption{\label{ta1} Parameters for DSA and SSA models: proton}
\begin{tabular*}{\textwidth}{@{}llrrrrlrlr@{}}
\hline
Regime&E[GeV]&$\sigma$&$\gamma$&$R_{max}$[GV]&$\mu$&$\mu_k$\\
\hline
DSA&0.5-330&1&2.50$\pm$0.03&50&2.85&3.85\\
SSA&$330-10^4$&1.16&2.35$\pm$0.03&$6\times 10^4$&2.70&4.2\\
\hline
\end{tabular*}
\end{minipage}
\end{table} 
\begin{table}[!ht]
\centering
\begin{minipage}{86mm}
\caption{\label{ta2} Parameters for DSA and SSA models: helium}
\begin{tabular*}{\textwidth}{@{}llrrrrlrlr@{}}
\hline
Regime&E[GeV]&$\sigma$&$\gamma$&$R_{max}$[GV]&$\mu$&$\mu_k$\\
\hline
DSA&0.5-245&1&2.42$\pm$ 0.03&$10^2$&2.75&3.1\\
SSA&$245-10^4$&1.15&2.30$\pm$ 0.03&$5\times 10^4$&2.65&4.5\\
\hline
\end{tabular*}
\end{minipage}
\end{table} 

To fit the observed data, we use the Zatsepin-Sokolskaya cosmic-ray model (\citealt{zatespin06}). In this model, a superposition of multiple types of sources reconstructs the proton and helium nuclei spectra, which vary with rigidity ($R$) according to:
\begin{equation}
Q(R)\sim R^{-\gamma}\times \phi (R)
\end{equation}
where $\phi (R)$ characterizes transition of spectral index from effective acceleration region for each type of sources to the after termination of the effective acceleration. This term is given by: 
\begin{equation}
\phi (R)=[1+(R/R_{max})^2]^{(\mu - \mu_k)/2}
\end{equation}
where $\mu=\gamma + \delta$ and $\mu_k$ is the spectral index after termination of effective acceleration and $R_{max}$ is the \textit{termination rigidity}. In order to reconstruct the particle spectra, we should convert the rigidity spectra to energy spectra:
\begin{equation}
R=\frac{1}{Z}\times \sqrt{E^2+2m_p\times A\times E},
\end{equation}
where $Z$, $A$ and $m_p$  are particle charge, atomic weight and proton mass, respectively. DSA and SSA inject particles with different spectral indices, as would be assumed different sources, so this model can be applicable in our study.

\begin{figure}
 \centering
 \includegraphics[width=3.4in]{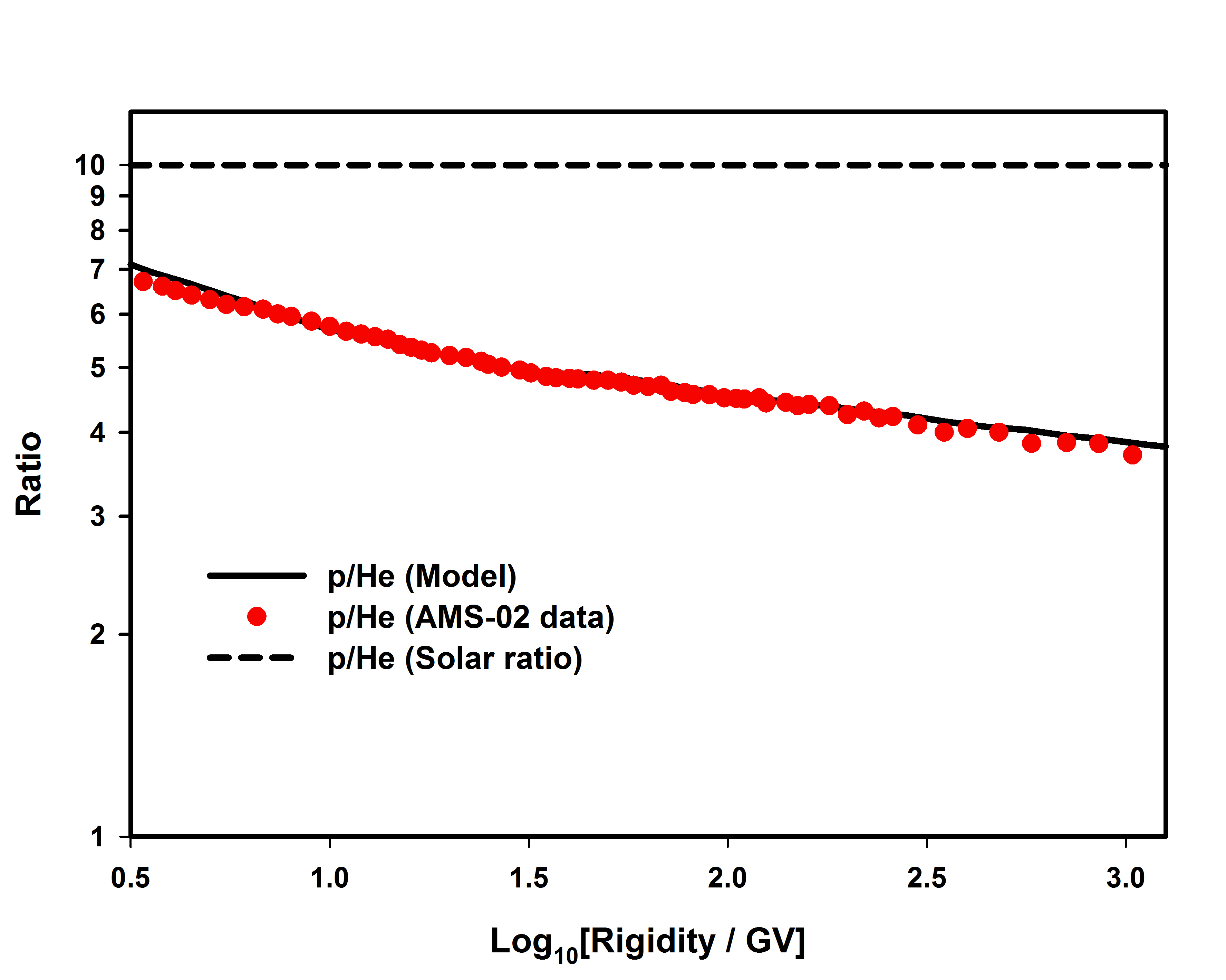}
 \caption{The proton to helium ratio (p/He) as a function of rigidity $R$.}
 \label{fig_ratio}
\end{figure}
Recent measurements of proton and helium nuclei spectra beyond the heliosphere by Voyager (\citealt{stone13}), have opened a new window for our understanding of the effects of solar modulation. For this purpose, we applied the solar modulation effect to reconstruct the spectra of these data as (\citealt{boezio97}): 
\begin{equation}
\rm{flux}_{mod}(E)=\rm{flux}(E+Ze\times \Phi)\times P,
\end{equation}
\begin{equation}
P=\frac{E^2+2m_pE}{(E+Ze\times \Phi)^2+2m_p(E+Ze\times \Phi)},
\end{equation}
where $\Phi$ is the modulation parameter and $e$ is the electron charge. We assume $\Phi =550$ MV. 

Figure \ref{fig_proton} shows our results on reproducing the observed proton spectrum by several CR detectors. In this figure, we show that the AMS-02 and CREAM experiments have observed anomalous  CR transport for protons with $E>330$ GeV. In particular, precise measurements by AMS-02 indicate a transition from diffusive to superdiffusive shock acceleration. We list the parameters used to calculate the fluxes in Tables \ref{ta1} and \ref{ta2}. The power law indexes, which were left as free parameters, agree with the values measured by AMS-02 experiment mentioned in \S. 2.

\section{Conclusion and discussion}
In this paper, we argued that a transition from DSA to SSA in SNRs is an acceptable explanation for the recent results of AMS-02, especially the proton and helium hardenings at $\sim 300$ \rm{GV}.  We proposed that the observed break in proton and helium spectra originates from the source due to different acceleration mechanisms and finally with different injection forms. 

There are theoretical explanations for the observed spectral breaks in CRs such as the models that interpret the breaks as a result of CR propagation in the ISM (\citealt{blasi12,tomassetti12}) while there are other models  explaining the mechanisms to produce the breaks at the sources (\citealt{zatespin06}). The latter scenario is widely adopted by some phenomenological works (\citealt{feng16}). It has been  proposed also that in addition to the Galactic sources, the local sources may affect the CR spectra due to spectral difference between them (\citealt{kawanaka11}).

The spectra of injected particles from the sources are distorted due to solar modulation and CR diffusion issues such as their interactions with ISM and spallation\footnote{In the literature, this effect is considered for primary particles like He and heavier primary nuclei, e.g., carbon nuclei.} while travelling the path to Earth. However, the spallation of helium is not so effective as to change the spectrum \citep{vladimirov12}. The effect of CR interaction with the ISM can be estimated from the B/C ratio, which we have used in this work, in addition to applying solar modulation to reconstruct measured spectra.

Obviously, the precise AMS-02 measurements show that spectral indices of proton and helium are different and the helium index is harder than that of the proton. As discussed widely by \citealt{ohira16}, this evidence can be explained by four different models: propagation, different sources, injection and the inhomogeneous environment. According to the last model, non-uniform environments produce the CR species with different spectra. As we showed in the last section, CR helium is produced in stronger shocks than those of protons. Since in the inner regions of shock, helium is more abundant and the shock is stronger (\citealt{ohira16}), CR helium and proton may originate from different parts of the same source, given significant inhomogeneity. 

Both of the proton and helium spectra show a pronounced change of slope around 300 GeV/n, which we interpret as the region where the DSA becomes less important than the SSA. Dependency of diffusion regime on energy is not determined yet, but it has been demonstrated that the divergence of magnetic field lines and  particle separation can only exist on scales less than the injection length scale of turbulence for the superdiffusive regime (\citealt{lazarianyan}). However, \cite{zimbardo13} showed that the diffusion coefficient is a function of particle speed, characteristic length scale and $\sigma$. 

Fig.\ref{fig_ratio} shows that the He/p ratio decreases with a single power law and indicates that solar modulation and interaction of He nuclei with ISM reduces the flux at lower energies. Furthermore, this figure shows that the abundances of the CR helium are larger than the solar abundance (see \citealt{lodders03}).

In addition, this model can be used for recent observed carbon spectrum by the AMS-02 experiment which is similar to that of helium, and 
  for heavier nuclei, which will be provided in the near future. However, in the case of the observed break for the secondaries such as lithium nuclei, this model cannot be used because the secondaries are not source-dependent and they are produced during propagation of primaries or at the local sources which are far from the scope of this model.

Moreover, in the higher energies, the knee observed in the cosmic hydrogen and helium spectrum by indirect measurements like ARGO-YBJ (\citealt{bartoli15}) at the energy $\sim$ 1 PeV suggests that our model needs additional components. This will be discussed in future papers.

\section*{Acknowledgements}
This work has been partially supported by research grant from the Ministry of Science and Technology (MoST) of Taiwan (\# 105-2112-M-008). We also acknowledge useful discussions with Prof. Shih-Chang Lee (Academia Sinica, Taiwan), Prof. Yuan-Hann Chang (National central University, Taiwan) and Matthew Behlmann (MIT, USA).



\label{lastpage}

\end{document}